\documentclass[a4paper, english,11pt]{article}
\usepackage{color}
\usepackage{dirtree}
\usepackage[utf8]{inputenc}
\usepackage[T1]{fontenc}
\usepackage{wrapfig}
\usepackage{float}
\usepackage[usenames, dvipsnames]{xcolor}
\usepackage{amsmath}
\usepackage{amssymb}
\usepackage{mathrsfs}
\usepackage{makeidx}
\usepackage{enumitem}
\usepackage{tabularx}
\usepackage{stmaryrd}
\usepackage{mathtools}
\usepackage{bm}
\usepackage{import}
\usepackage[left=2.5cm,right=2.5cm,bottom=3cm]{geometry}
\usepackage{graphicx}
\usepackage{makecell}
\usepackage{fancyhdr} 
\usepackage{comment}
\usepackage{indentfirst}
\usepackage{verbatim}
\usepackage[table]{colortbl}
\usepackage{pgfplotstable}
\usepackage[toc,page]{appendix}
\usepackage{nicematrix}
\usepackage{amsmath,amssymb}
\NiceMatrixOptions{nullify-dots}
\usepackage{svg}
\usepackage{subcaption}
\usepackage[numbers]{natbib}
\usepackage{authblk}
\usepackage{gensymb}
\usepackage{tikz}
\usetikzlibrary{patterns}
\usetikzlibrary{shapes}
\usetikzlibrary{arrows.meta}
\usetikzlibrary{calc,shadows,matrix}
\usetikzlibrary{decorations.pathreplacing}
\usetikzlibrary{external}
\usepackage{pgfplots}
\usepgfplotslibrary{patchplots}
\usepgfplotslibrary{fillbetween}
\usepgfplotslibrary{polar} 
\usetikzlibrary{pgfplots.polar}
\usepgfplotslibrary{colormaps}
\title{Practical implementation of a chiral phononic crystal demonstrator with ultra-low frequency bandgap}
\date{\vspace{-5ex}}
\author[1,2]{L. Mardini}
\author[1]{A. Bergamini}
\author[3,4]{C. Claeys}
\author[2,4]{E. Deckers}
\author[1]{B. Van Damme}

\affil[1]	{Empa, Laboratory for Acoustics/Noise Control, Ueberlandstrasse 129, 8600 Dübendorf, Switzerland}

\affil[2]	{Department of Mechanical Engineering campus Diepenbeek, KU Leuven
Wetenschapspark 27, B-3590, Diepenbeek, Belgium}

\affil[3]		{Department of Mechanical Engineering campus Heverlee, KU Leuven, Celestijnenlaan 300, B-3001, Heverlee, Belgium }

\affil[4]		{Flanders Make@KU Leuven, Belgium }
\pgfplotsset{compat=1.18} 
\begin{document}

\maketitle 
\date{}

\section*{Abstract}

The use of phononic crystals for vibration attenuation and isolation has been widely studied, showing that the attenuation frequency range depends on their mass and stiffness.
The concepts of chirality and tacticity have been introduced into classical phononic crystals to enrich the dynamics of the mass elements and thereby achieve lower frequency  ranges with high vibration attenuation. Although these concepts have demonstrated their effectiveness on lab-scale crystals, their implementation in industrial applications is still rare. 
Chiral phononic crystals require a complex geometry that complicates their manufacturing. Existing examples require to be fabricated by 3D printing, making them expensive to build on a large scale for demonstration purposes or in-situ applications. In this study, we redefine a chiral phononic crystal design for translational-rotational coupling in order to enable its manufacturability using exclusively conventional processes. We then investigate the design space of these newly designed phononic crystals, using a simplified unit cell FEM model that minimizes computation time. A parametric study is conducted to investigate the crystal's tunability by modifying the dimensions of the chiral links between the masses. A large crystal with ultra-low frequency range attenuation -- starting at 60~Hz -- is then designed, with the aim to demonstrate the influence of the crystal's tacticity on the vibration isolation by hand sensing. A crystal composed of 2 unit cells is manufactured and its measured transfer function is compared with numerical predictions, thus highlighting the disparities between the behavior of the structure under real-life and ideal excitation conditions.

\newpage
\section{Introduction}

Phononic crystals, a periodic arrangement of masses - the atoms -  interconnected by springs - the bonds -, represent a widely investigated concept for vibration isolation. In periodic structures, two mechanisms can be distinguished at the origin of bandgap generation. A Bragg bandgap observed in phononic crystals is due to Bragg scattering and therefore to lattice periodicity \citep{ouakka_efficient_2023,hedayatrasa_optimization_2018,meseguer_rayleigh-wave_1999}, whereas a local resonance bandgap as observed in resonant periodic structures only depends on the modal properties of the resonators \citep{liu_locally_2000,ruan_isolating_2021}. 
The bandgap generation often results from a combination of these two phenomena, meaning that the Bragg bandgap depends on the internal resonance of a unit cell \citep{liu_wave_2012,tian_merging_2021,matlack_composite_2016}. The coupling of these two mechanisms enables the two types of bandgaps to be merged and the attenuated frequency range to be broadened. However, attenuating low-frequency vibrations remains a challenge that is bounded by real-life constraints.  Indeed, in a monoatomic phononic crystal waves can propagate up to a frequency of $f=\frac{1}{2\pi}\sqrt\frac{4k}{m}$ with $k$ the stiffness between two contiguous atoms and $m$ the mass of the atoms \citep{brillouin_wave_1946,deymier_acoustic_2013}. To shift the resonance frequency to a lower frequency, either the atom's dynamic mass has to be increased or the spring's dynamic stiffness has to be reduced. 
These two principles conflict with two typical engineering constraints. Every practical application requires a minimum axial static stiffness to sustain the attenuated structure. On the other hand, the static mass cannot be infinitely increased.  The challenge is therefore to decouple the static and dynamic properties of the crystal.  

\begin{figure*}
    \centering
\includegraphics[width=0.5\textwidth]{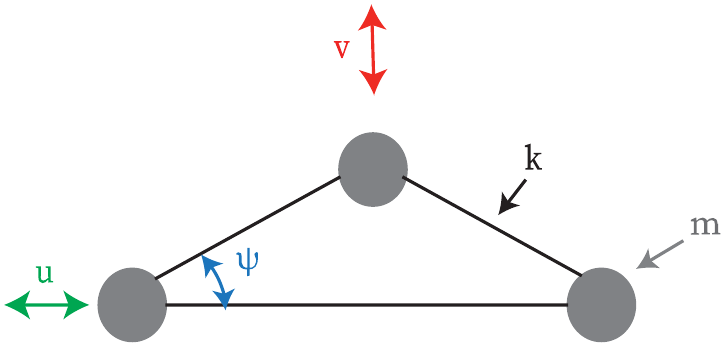}
    \caption{Inertia amplification mechanism proposed in \cite{yilmaz_phononic_2007}.}
    \label{fig:Yilmaz}
\end{figure*}

\begin{figure}
    \centering
\includegraphics[width=0.5\textwidth]{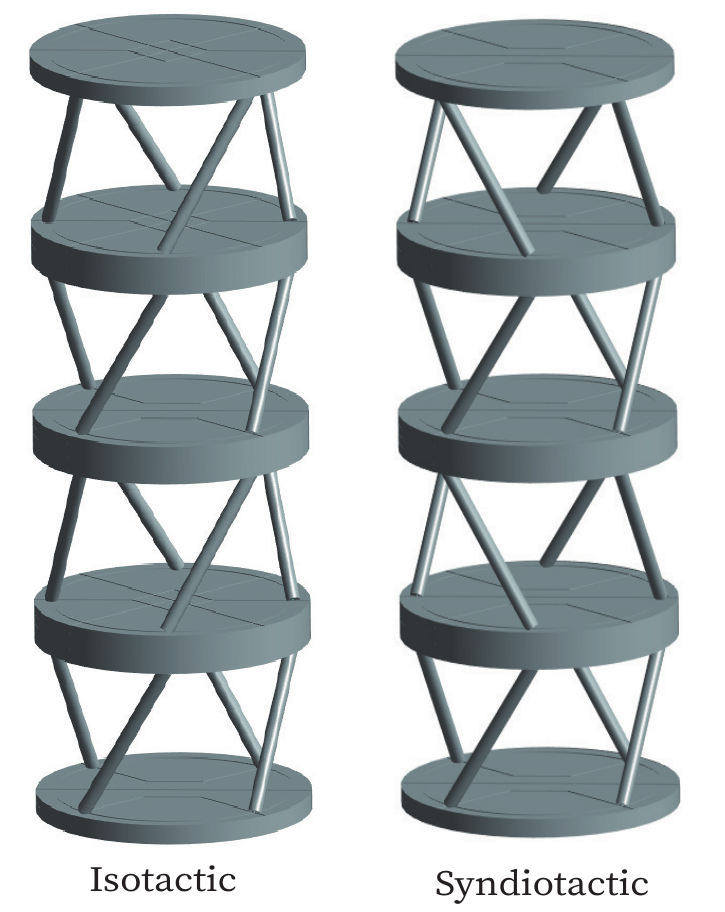}
    \caption{Isotactic (left) and syndiotactic (right) orientations of the chiral links connecting the disks of a chiral phononic crystal.}
    \label{fig:Tacticity}
\end{figure}

To overcome the limitations described above,\citet{yilmaz_phononic_2007} proposed to inertially amplify the masses of the chain. In this paper, two consecutive masses are no longer connected by stiff beams along the same axis as shown in Figure~\ref{fig:inertiaAmpli}~(a). Offsetting the auxiliary masses from the main axis creates an angle  $\psi$, and yields a combination of their translations in two directions, which increases the effective mass along the wave propagation direction. Therefore, a motion $u$ of one mass of the chain along the main axis induces a motion $v$ of the next mass perpendicular to this axis with the relation $v=\frac{u}{\tan\psi}$ as represented in Figure~\ref{fig:Yilmaz}. This coupling mechanism allows the starting bandgap frequency of a one dimensional phononic crystal to be shifted to a lower frequency while maintaining the static mass of the atoms and static stiffness of the springs of the chain and introducing only a minor mass penalty into the unit cell. This idea can also be applied to chiral phononic crystal as proposed in \citep{orta_inertial_2019,krushynska_accordion-like_2018,delpero_inertia_nodate}. Here, tilted connectors link the finite size masses of a crystal to introduce a new coupling mechanism in these crystals. Similar to the translational coupling in \cite{yilmaz_phononic_2007}, a translation $u$ of a mass induces its rotation due to kinematic constraints. As shown in \cite{ding_description_2023}, for a given axial stiffness and quasi equal density, a chiral crystal exhibits a lower frequency bandgap than a non-chiral crystal. For a finite number of parameters such as the connectors' positioning, their inclination angle, and their shape as well as the atoms' mass and rotational inertia, these crystals ensure a certain tunability of the amplified inertia and the stiffness of the structure. However, the relative orientation of the chiral links in subsequent cells, known as tacticity, greatly influences the propagation of waves  \cite{bergamini_tacticity_2019}. If the tilting connectors are oriented in opposite directions on either side of the mass, the arrangement is said to be syndiotactic as shown in Figure~\ref{fig:Tacticity}. If the orientation is the same on either side of the mass, the arrangement is called isotactic. In \cite{bergamini_tacticity_2019} it is shown that a syndiotactic arrangement achieves a wide full bandgap whereas an isotactic arrangement can be an efficient isolator for longitudinal wave propagation, but it does not attenuate flexural waves in the same frequency range.


These theoretical concepts have been demonstrated and proved their effectiveness on lab-scale crystals, but they remain rare in industrial applications. To illustrate the advantages of using a chiral crystal rather than a conventional vibration damper, a demonstrator is an effective way of bridging the gap between theory and practice to convince potential users, especially if the bandgap can be felt by the user without the need for measuring equipment. However, building large-scale chiral phononic crystals enabling to illustrate the abstract concepts of chirality and the influence of their tacticity presents several challenges. First, the chiral phononic crystal  requires to attenuate vibrations in the ultra low frequency range (below 150 Hz \cite{oroszi_vibration_2020}), enabling the human senses to feel the crystal's bandgap by hand. Secondly, the design of the crystal needs to be redefined so that a large sample can be manufactured, avoiding 3D printing but using conventional processes. Thirdly, this new complex design requires fast numerical modeling to perform parametric studies and facilitate its dimensioning.

In this paper, improved designs of isotactic and syndiotactic crystals are studied, limiting the number of mechanical components and machining operations. Exclusively conventional processes such as turning, milling, laser cutting or folding processes are used for their production. This work highlights the geometric constraints brought by this new design. Taking these geometrical constraints into account, a FEM model, accurately approximating the geometry but with a small mesh size, is built. This parametric model is used to investigate the tunability of the crystal's bandgaps by performing a unit-cell analysis to calculate the dispersion diagrams. 

By employing the direct wave finite element method, the complex wave number as a function of frequency allows the detailed investigation of the efficiency of the isolation properties, including complex effects such as the dynamic deformation of the crystal's building blocks. Additionally, harmonic response simulations of a finite-size sample of the chosen design are used to investigate non-ideal excitation conditions that might affect the identified bandgap of a finite sample of the phononic crystal. 

The rest of this paper is structured as follows. Section \ref{Dimensioning} describes the dynamics governing chiral phononic crystals and a new geometry which preserves these physical principles while improving manufacturability. In Section \ref{ParamStudy}, the FEM approximated geometry is described and a parametric study of this new design is presented. The crystal's dispersion curves are numerically calculated based on the unit cell FEM model. Section \ref{ExpNum} describes the experimental and numerical setup of a syndiotactic and isotactic chiral phononic crystal consisting of only two unit cells. The measured and calculated transfer functions are compared and discussed. 

\section{Dimensioning and practical design of phononic crystals with inertia amplification}\label{Dimensioning}

\subsection{Concept of translational-rotational inertia amplification}

\begin{figure}[hbt!]
\centering
\includegraphics[width=0.5\textwidth]{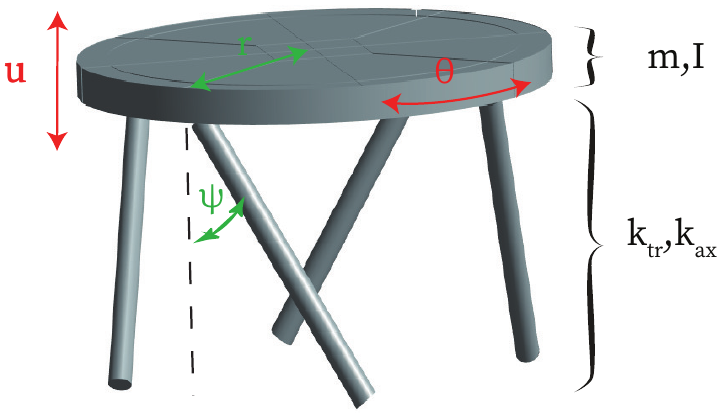}
        \caption{Schematic representation of a translational-rotational oscillator \cite{bergamini_tacticity_2019}, with four chiral links. The tilting connectors form an angle $\psi$ with the vertical axis and are joined to the outer edge of the masses represented by disks. These links couple the translation $u$ to the rotation $\theta$ of the mass.}\label{fig:inertiaAmpli} 
\end{figure}

Phononic crystals for elastic wave propagation can be defined as periodic arrangements of masses $m$ connected by springs of stiffness $k$ \citep{brillouin_wave_1946,deymier_acoustic_2013}. In a one-dimensional crystal, the motion of the masses in the chain is described by a single translational degree of freedom $u$. By introducing stiff hinged chiral links between the masses as shown in Figure~\ref{fig:inertiaAmpli}~(b), the translation is coupled to the rotation $\theta$ of the masses by the relationship $\theta=\frac{u}{r\tan\psi}$ with $r$ the radius of eccentricity and $\psi$ the inclination angle of the struts. A single translational-rotational oscillator (TRO), which can be used as the base to build a chiral chain \citep{orta_inertial_2019,delpero_inertia_nodate,bergamini_tacticity_2019} or as local resonator to form a metamaterial by periodically placing them on a host structure \citep{zhao_nonlinear_2024, van_damme_inherent_2021}, exhibits the following equation of motion

\begin{equation}
    F= \tilde{M}\ \ddot{u} + \tilde{C}\dot{u} \vert u\vert +\tilde{K}u 
\end{equation}
with 

\begin{equation}
    \tilde{M}=m+\frac{I}{r^2\tan^2\psi},
\end{equation}

\begin{equation}
    \tilde{K}=k_{lax}+\frac{k_{tr}}{r^2\tan^2\psi},
\end{equation}

\begin{equation}
    \tilde{C}=\frac{I}{Lr^2\sin^2\psi},
\end{equation}
and $k_{ax}$ and $k_{tr}$ are the axial and torsional stiffness, respectively, coming from the chiral links with length $L$. $I$ is the moment of inertia of the disk.

In an ideal one-dimensional phononic crystal, the bandgap opens at twice the first mass-spring resonance. From this equation, it can be inferred that the resonance frequency $f=\frac{1}{2\pi}\sqrt{\frac{\tilde{K}}{ \tilde{M}}}$ of the TRO depends on the axial and torsional stiffness, on the angle of inclination and radius of eccentricity of the chiral links as well as the mass and moment of inertia of the atom. To minimize the resonance frequency, the objective is twofold: to maximize $\tilde{M}$ and minimize $\tilde{K}$. In order to increase $\tilde{M}$, we can take advantage of the inertia amplification term. For a crystal with a low weight, a geometry with a high moment of inertia must be designed. To maximize the moment of inertia, the mass requires to be concentrated away from the axis of rotation. Minimizing $\tilde{K}$, on the other hand, involves a more complex design. Indeed, the axial and rotational stiffness are coupled by the shape of the chiral links, their connection to the disks, and their angle of inclination. Due to the intricate details, these two stiffness values can only be calculated from finite element simulations. If the model is sufficiently fast, a parametric study can be performed to optimize the shape of the chiral links to reach a targeted bandgap.

\subsection{Design for Conventional Manufacturing}
The introduction of chiral links to lower the bandgap frequency not only complicates crystal tunability, but also its manufacturing process. For this reason, chiral structures are often 3D printed \citep{jiao_design_2021,bergamini_tacticity_2019}. Few studies have considered the fabrication of chiral structures using conventional processes. To our knowledge, only 2D planar structures have been manufactured using exclusively conventional processes such as water-jet cutting as in \cite{zhu_chiral_2014} and 3D structures have been fabricated combining laser cutting with 3D printing as in \cite{orta_inertial_2019}. In order to be able to create large demonstrator crystals, its design is here rethought to respond to the challenge of minimizing the number of assembly steps, and to allow the complex shape to be entirely made by conventional manufacturing processes. 

\begin{figure}[H]
\centering
\includegraphics[width=\textwidth]{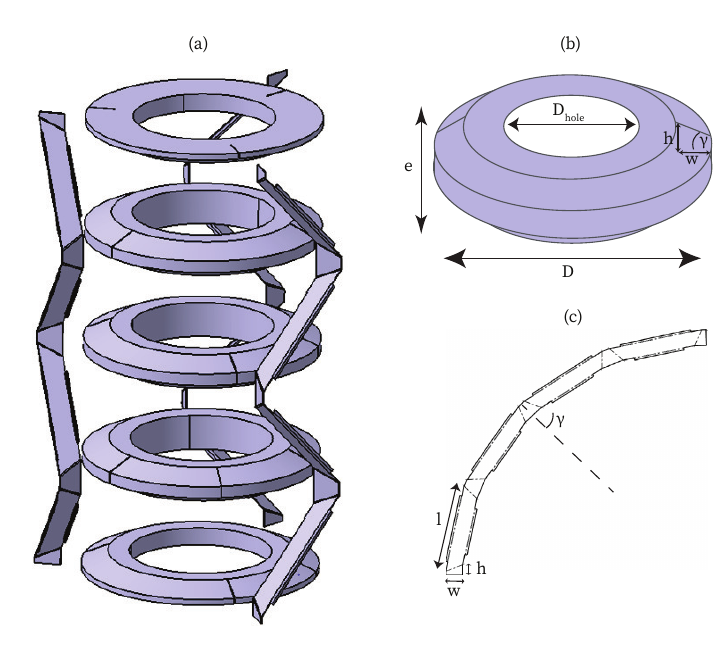}
            \caption{Components and assembly of crystal elements. (a) Exploded view of the assembly.  (b) Iso view of the manufactured ring. (c) Construction drawing of the unfolded strut.} \label{fig:Manufacturing} 
\end{figure}

Previous studies have shown that two unit cells of the syndiotactic configuration are sufficient to achieve high isolation capability \cite{bergamini_tacticity_2019}. The phononic crystal design proposed here contains five masses, shown in Figure~\ref{fig:Manufacturing}~(b), interconnected by twelve tilted struts. Instead of manufacturing the connectors individually, only one single mechanical component is needed to form four connectors, thus reducing the number of assembly steps as shown in Figure~\ref{fig:Manufacturing}~(c). When accurately designed, the thin metal parts can be folded to achieve the desired inclination angle as shown in Figure~\ref{fig:Manufacturing}~(a). To maximize the moment of inertia of the atoms while keeping their weight low, a ring shape is used. Thereafter, the mechanical components constituting the struts are inserted into the masses to form the crystal as shown in Figure~\ref{fig:Manufacturing}~(a). 

\begin{figure}[hbt!]
     \centering
\includegraphics[width=0.5\textwidth]{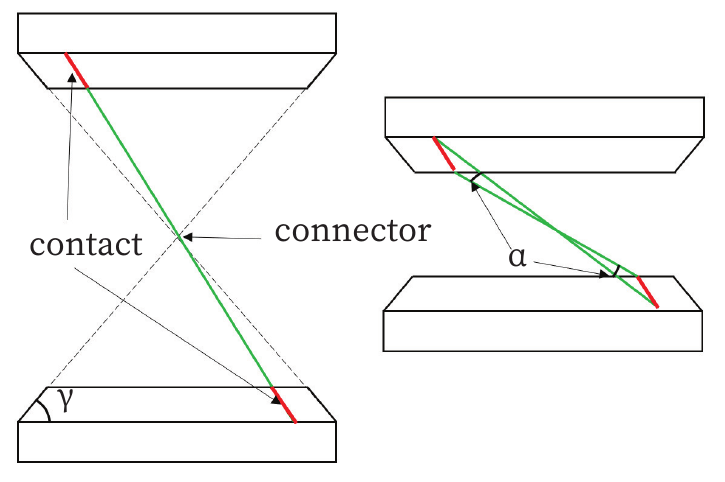}
     \caption{Schematic representation of the required alignment between cone tips (left) to avoid the introduction of a twist angle $\alpha$ of the struts (right).}
     \label{fig:TwistingAngle}
 \end{figure}

However, this proposed assembly needs to respect several geometric conditions. To ensure that the struts tilt without torsional deformation, they must be radially in contact with the rings. This condition cannot be met by giving a cylindrical ring shape to the masses without twisting the struts between the contact points. To avoid this unwanted deformation, we propose to modify the shape of the masses. By chamfering their bottom and top outer edges, two conical surfaces with their vertices on the axis halfway between the masses are created to ensure radial contact as shown in Figure~\ref{fig:TwistingAngle}~(b). The angle $\gamma$ formed by this cone depends on the distance between the masses, and their diameter. The cone's tip on each mass must be aligned to avoid the introduction of a twisting angle $\alpha$ as shown in Figure~\ref{fig:TwistingAngle}~(b). 

Finally, to avoid bending modes of the connectors at low frequencies, a c-shaped cross-section is used to stiffen the struts. Nevertheless, this cross-section is particularly problematic at the joints with the masses, which should act as hinges and require a low bending stiffness. Thus, close to the connections, the struts are not stiffened and shaped with a rectangular cross-section. This innovative design enables an assembly in a reduced number of steps: after folding, the struts are inserted into the rings and glued at their joints.

The individual parts of the crystal can be manufactured by conventional techniques. Indeed, the masses designed can now be fabricated using turning and milling processes. The diameter $D_{hole}$ milled allows the mass and moment of inertia of the rings to be tuned. The struts are manufactured by laser cutting and folding processes. A plate of thickness $t$ is laser cut to form the profile in Figure~\ref{fig:Manufacturing}~(c) constrained by the conical shape of the cone and its thickness. Thereafter, depending on the chosen tacticity, the cut geometry is folded. For the isotactic crystal, the folding process is performed invariably in the same direction.  In contrast, for the syndiotactic crystal, the plate is alternately folded in two directions.
 
 
\section{Parametric study of the dispersion characteristics based on a unit cell model}\label{ParamStudy}

\subsection{Dispersion relation of isotactic and syndiotactic phononic crystals}
In this section, we investigate the crystal's dynamic response as a function of its dimensions, taking into consideration the previously defined geometry. More specifically, we are examining the influence of the shape of the chiral connectors on the starting frequency of the longitudinal bandgap. In order to clearly demonstrate the concept of vibration isolation, this frequency is tuned at 60~Hz. This is sufficiently low to sense a variation in vibrational amplitude by hand, so that the difference in isolation efficiency between different tacticities can be illustrated. As the aim is to build a movable demonstrator, it must have a moderate weight (around 5~kg per crystal) and a compact size (height of 40~cm).\\ 

By applying the Bloch-Floquet periodic boundary condition to a unit cell FEM Model, the Wave Finite Element method can be used for computing the dispersion curve of 1D \cite{mace_finite_2005} %or 2D \cite{mace_modelling_2008}
infinite periodic structures. In a one-dimensional periodic structure, the degrees of freedom on one side of the unit cell are related to those on the opposite side by a phase shift yielded by the so-called Bloch wave. This constraint eliminates the degrees of freedom on the so-called slave side of the unit cell, thus reducing the dynamic stiffness matrix. The eigenvalue problem is further reduced by eliminating the internal nodes not involved in the periodicity using dynamic condensation. Solving the quadratic eigenvalue problem using a direct approach \cite{hussein_dynamics_2014,van_belle_impact_2017}, the complex wavenumber is calculated as a function of frequency, and provides information on whether the wave is propagating, evanescent or decaying \cite{cool_guide_2024}. A purely real wavenumber indicates a propagating wave, a purely imaginary wavenumber an evanescent wave and a complex wavenumber a decaying wave.

\begin{figure}[h]
       \centering 
\includegraphics[width=\textwidth]{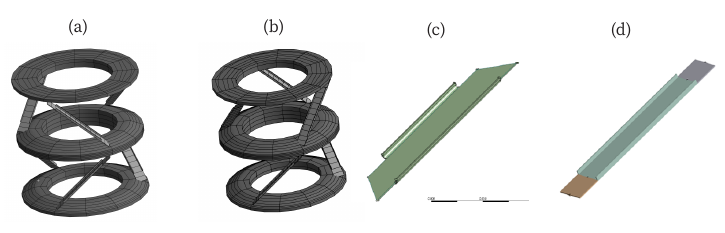}
        \caption{Meshed unit cell models in syndiotactic (a) and isotactic (b) arrangements. (c) shows the real geometry of the manufactured connectors. The approximated geometry (d) reflects their real bending behavior by assigning cross sections to beam elements.}\label{fig:FEMSyndioIso} 
\end{figure}

The unit cell FE model shown in Figure~\ref{fig:FEMSyndioIso}~(a) and (b)  simplifies the crystal's geometry to minimize computation time. The rings are modelled as elastic solids, interconnected by three struts discretised  by beam elements.  For simplification purposes, the strut's geometry shown in Figure~\ref{fig:FEMSyndioIso}~(c) is approximated by dividing the strut into three parts as shown in Figure~\ref{fig:FEMSyndioIso}~(d).  Near the joint with the rings a rectangular cross-section is applied while in the center a c-shaped cross-section is assigned. The isotactic arrangement presents struts with identical orientation as shown in Figure~\ref{fig:FEMSyndioIso}~(b), while the syndiotactic arrangement shows struts with alternating orientation as presented in Figure~\ref{fig:FEMSyndioIso}~(a). 

The crystal is initially given the dimensions shown in Table~\ref{Tab:Dimensions}. Aluminium is used for the rings, and steel for the struts, with values given in Table~\ref{Tab:MaterialProperties} 
\begin{table}[hbt!]
\caption{Crystal's dimensions.}
\begin{center}
        \begin{tabular}{ |c|c|c|c|c|c|c|c| } 
        \hline
        $D_{hole}$ & $D$ & $e$ & $h$ & $w$ & $t$ & $l$ & $\psi$\\ 
        \hline
        \hline
        18 cm  & 20 cm & 2.775 cm & 0.8875cm & 2 cm & 0.07 cm & 12.117 cm & 48\degree \\ 
        \hline
    \end{tabular}
\end{center}
\label{Tab:Dimensions}
\end{table}

\begin{table}[hbt!]
\caption{Material properties}
\begin{center}
        \begin{tabular}{ |c|c|c|c| } 
        \hline
         & Density $\mathrm{kg.m^{-3}}$ & Young Modulus (GPa) & Poisson ratio \\ 
        \hline
        \hline
        Aluminium & 2770 & 71 & 0.33  \\ 
        \hline
        Steel & 7850 & 200 & 0.3  \\ 
        \hline
    \end{tabular}
\end{center}
\label{Tab:MaterialProperties}
\end{table}

\begin{figure}[hbt!]
    \centering 
    %\externalremake
\includegraphics[width=\textwidth]{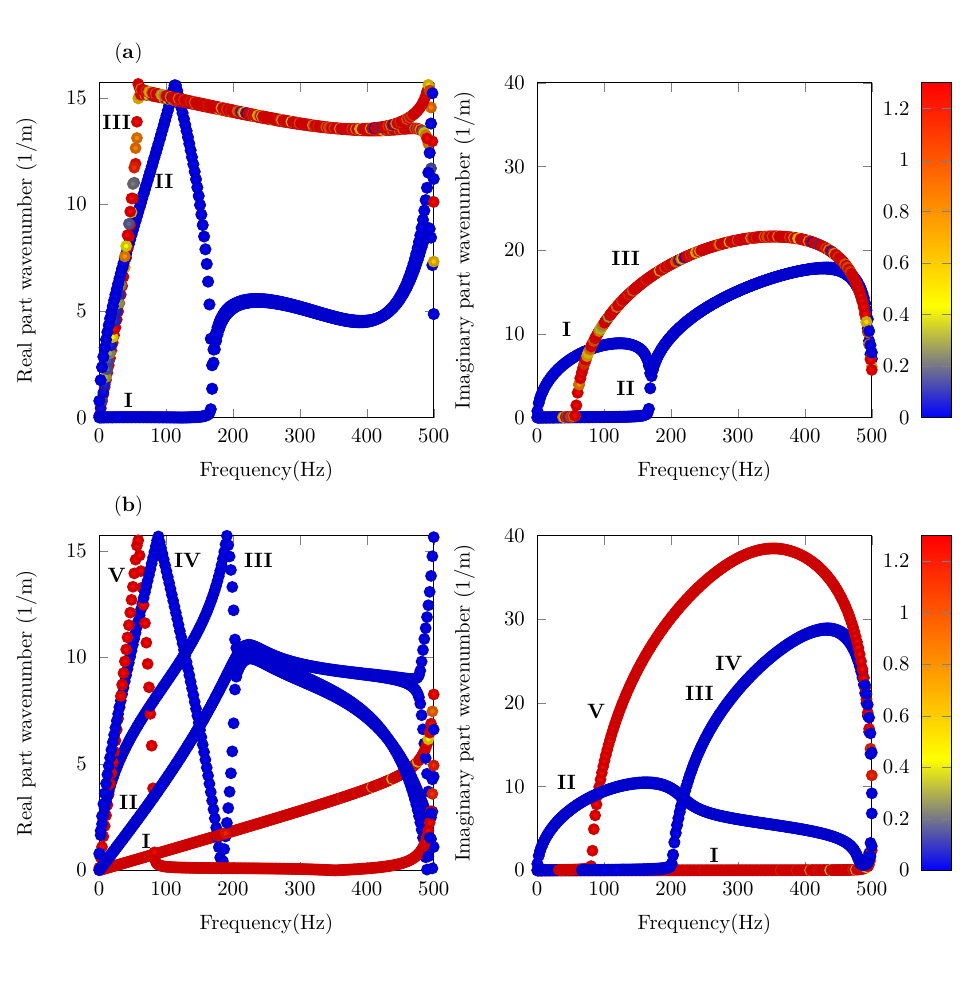}
    \caption{Dispersion curves of the syndiotactic (a) and isotactic (b) crystal showing real part's (left) and imaginary part's (right)  wavenumber  (1/m) with respect to frequency (Hz).}
    \label{fig:SYNDIODispersion}
\end{figure}

The dispersion relations of isotactic and syndiotactic chiral arrangements are first compared  within the frequency range 0-500Hz (Figure~\ref{fig:SYNDIODispersion}). The dispersion curves are color-coded as function of the polarization of the atoms in the crystal. This value is calculated taking the absolute value of the average curl of the displacement field of the three rings in the direction of the principal axis, as proposed in \cite{bergamini_tacticity_2019}. A high polarization (red) indicates an important rotation component of the disks, while a low polarization (blue) indicates a unit cell deformation without rotational component. Since the disks' rotation is coupled to their translation, a high polarization implies a strong rotation/translation coupling and therefore demonstrates the presence of a longitudinal wave mode, whereas a low polarization indicates a weak coupling and rather the presence of a flexural or shear wave. As observed in the dispersion curve in Figure~\ref{fig:SYNDIODispersion}~(a), the syndiotactic crystal presents longitudinal (III) and flexural (I,II) waves within the frequency range considered. It can be seen from the dispersion curves that the longitudinal mode (III) and flexural mode (II) are propagative up to respectively ${f_{III}=56}$ Hz and $f_{II}=167$ Hz and then start to decay at these frequencies. The flexural mode (I) is not propagative since its wavenumber shows a large imaginary part over the whole frequency range.  Therefore, we can observe from the dispersion curves that the syndiotactic crystal achieves a longitudinal bandgap starting at 56 Hz and a full bandgap starting at 167 Hz. 

The isotactic crystal shows more complex dynamics as seen in Figure~\ref{fig:SYNDIODispersion}~(b). Longitudinal (I,V), bending (III,IV) and shear modes (II) are propagating within the frequency range considered. The longitudinal mode (V) and flexural modes (III,IV) are propagative up to respectively $f_{V}=80 Hz$ and $f_{III,IV}= 200 Hz$. At these 2 frequencies, partial longitudinal and flexural bandgaps could be expected. However, the longitudinal mode (I), showing a purely real wavenumber, is propagative over the entire range breaking the partial longitudinal bandgap identified. As shown in \cite{bergamini_tacticity_2019}, the syndiotactic crystal enables to achieve a low frequency full bandgap unlike the isotactic crystal which shows no full bandgap in the considered frequency range. For this reason, the following parametric study will focus on the syndiotactic arrangement.

\subsection{Parametric study of the syndiotactic crystal: geometry versus dispersion}
In order to find a suitable for the desired static and dynamic properties, a parametric study of the struts is performed. With respect to the exploration of the design space, we first investigate the influence of the struts' thickness and then the influence of their inclination angle on the dynamic behavior of the syndiotactic crystal. To this end, we compare the dispersion curves for 3 different thicknesses: 0.6mm, 0.7mm and 0.8mm. In a second step, we compare the behavior of the crystal for three angles: 43°, 49° and 55°. The other crystal dimensions remain as in Table~\ref{Tab:Dimensions}. 

\begin{figure}[hbt!]
    \centering
\includegraphics[width=\textwidth]{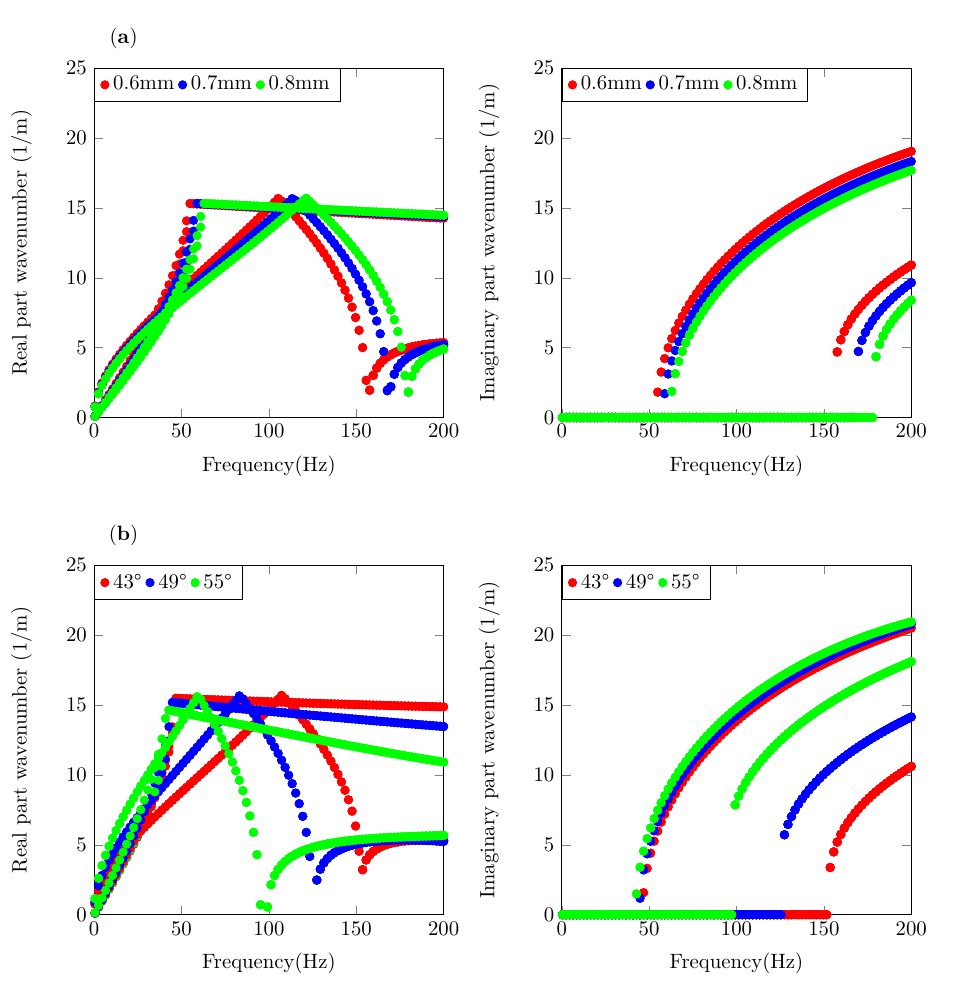}
    \caption{Dispersion curves of the syndiotactic crystal showing real part's (left)  and imaginary part's (right) wavenumber  (1/m) with respect to frequency (Hz). In (a) three strut's thickness are compared: in red the dispersion relation for a thickness of 0.6 mm, in blue  for a thickness of 0.7 mm and in green for a thickness of 0.8mm.  In (b), three strut's inclinations are considered: in red is plotted the dispersion relation for an angle of 43°, in blue for an angle of 49° and in green for an angle of 55°.}
    \label{fig:ThicknessDispersion}
\end{figure}

From the first comparison in Figure~\ref{fig:ThicknessDispersion}~(a), we can observe that increasing the thickness from 0.6 mm to 0.8 mm shifts
both bandgaps for the longitudinal and flexural modes by 15\% to higher frequencies.  The second comparison in Figure~\ref{fig:ThicknessDispersion} shows that an increase in the strut's inclination angle shifts the longitudinal partial bandgap and full bandgap starting frequency to a lower frequency. However, the trend between longitudinal and bending modes differs. Decreasing the inclination angle from 55° to 43° shifts the curve of the longitudinal mode by 9\% and the curve of the bending mode by 50\%. The angle of inclination has a greater impact on the bending mode of the crystal than on its longitudinal mode.

\begin{figure}[hbt!]
    \centering
\includegraphics{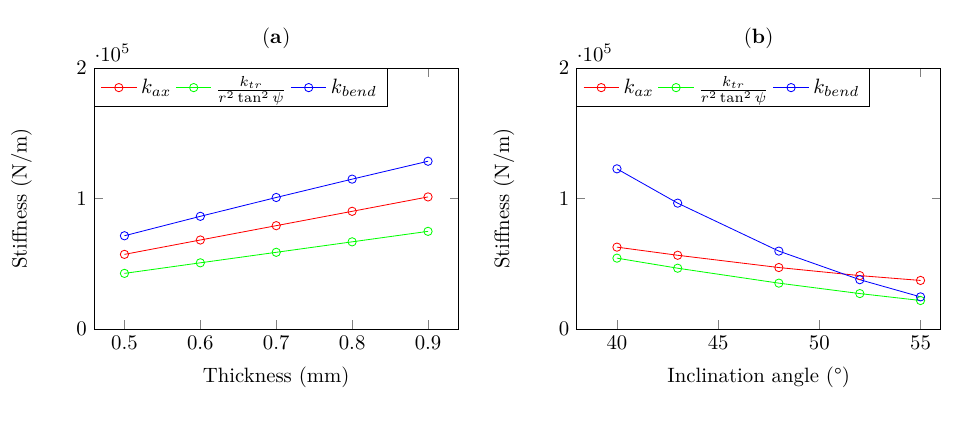}
    \caption{Axial (red), normalised torsional (green) and bending (blue) stiffness of the unit cell with respect to the thickness of the connectors (a) and their inclination angle (b).}
    \label{fig:AnglesStiffness}
\end{figure}

The unit cell's static properties are determined, in particular the axial, torsional, and bending stiffness, and compared with the observations made on the dispersion relation. The trends seen on the dispersion curve correlate with the conclusions drawn from a static analysis.  As observed in Figure~\ref{fig:AnglesStiffness}~(a), increasing the thickness of the struts increases the axial stiffness $k_{ax}$, normalized torsional stiffness $\frac{k_{tr}}{r^2\tan^2\psi}$ and bending stiffness  $k_{bend}$ of the crystal in a more or less linear way. It can be observed that the axial stiffness contributes twice more to the longitudinal rigidity of the structure than the normalized torsional stiffness. Increasing the thickness of the connectors stiffens the crystal evenly in axial and bending directions, which explains the equal shift of the starting frequency of the longitudinal bandgap and full bandgap observed on the dispersion curve. From this study, it can be inferred that the global properties of the crystal are not an explicit function of the local properties of the struts \cite{ding_origin_2024}, which explains the importance of a parametric study for its dimensioning. Indeed, the bending stiffness of the struts, unlike the global stiffness properties of the crystal, is a cubic function of their thickness. 

The inclination of the struts appears to be a major parameter to control the bending stiffness of the crystal while the axial and normalized torsional stiffness are less impacted by this parameter. As it can be seen on Figure~\ref{fig:AnglesStiffness}~(b), the axial and normalized torsional stiffness decrease in a similar trend while increasing the inclination angle of the struts as shown in Figure~\ref{fig:AnglesStiffness}. However, their rate of change is significantly lower than the one of the bending stiffness.

\section{Experimental and numerical validation of the optimized chiral crystal design.}\label{ExpNum}
\subsection{Experimental setup.}

\begin{figure}[hbt!]
        \centering
\includegraphics[width=\textwidth]{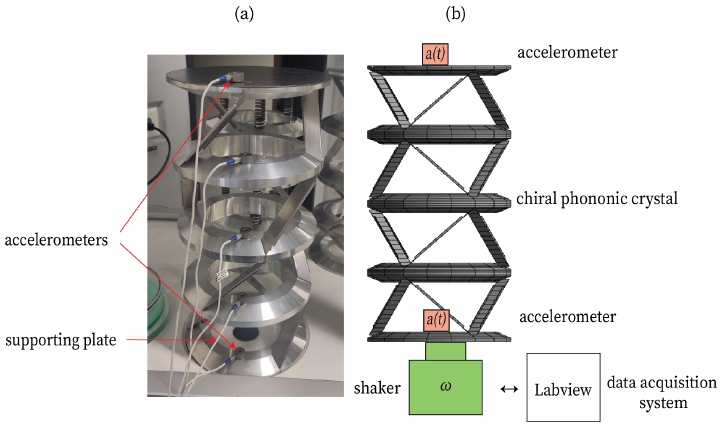}
        \caption{Picture (a) and schematic representation (b) of the experimental setup.}\label{fig:ExpSetUp}
\end{figure}

Following the previous parametric study, an isotactic and a syndiotactic phononic crystal are manufactured. The dimensions of the rings remain the same as indicated in the Table~\ref{Tab:Dimensions}. Each crystal measures 0.4m and weighs 5.432 kg.
The experimental set up is presented in Figure~\ref{fig:ExpSetUp}. The phononic crystal rests on a circular plastic support plate screwed to a shaker. The shaker, connected to an amplifier, vertically excites the support plate. The motion of the crystals' bottom and top ring can thus be sensed by hand for demonstration purposes. Using in-house coded Labview software, a sinusoidal sweep force is generated by a National Instruments PXI-5402 card and used to excite the bottom ring. The crystal's vibration isolation capacity is measured by two accelerometers (type ISOTRON 7251A): one on the top ring and one on the bottom ring. The signals are captured by a NI PXI-4496 data acquisition card. The transfer functions of the isotactic and syndiotactic crystals are defined as the ratio between the acceleration measured at the top and bottom. 

\subsection{Numerical setup.}

\begin{figure}[hbt!]
        \centering
        \includegraphics[width=0.5\textwidth]{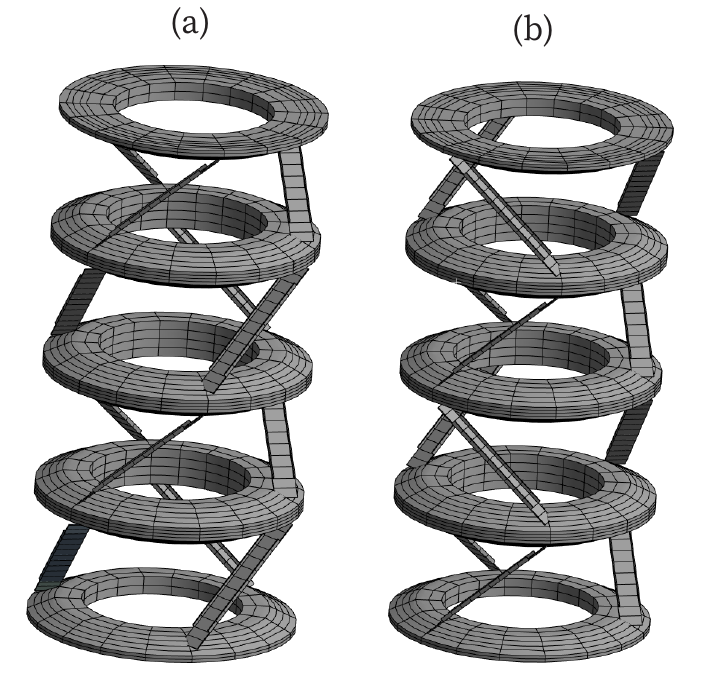}
        \caption{Meshed two unit cell models in isotactic (a) and syndiotactic (b) arrangements.}\label{fig:FullSyndioIso} 
\end{figure}

In this section, we model the finite-sized crystals composed of two unit cells in order to study the impact of sample size and edge effects on crystal behavior, which are not captured by dispersion analysis. The two crystals are built on the basis of their unit cell geometry described in Section~\ref{ParamStudy}.  In order to model the syndiotactic crystal shown in Figure~\ref{fig:FullSyndioIso}~(a), the unit cell is simply translated. To model the isotactic crystal shown in Figure~\ref{fig:FullSyndioIso}~(b), the unit cell is translated and additionally rotated over an angle of 60° so that the connecting struts align.
A harmonic displacement is imposed on the face of the bottom ring. Thereafter, the displacement on the face upper ring is retrieved. The transfer function is calculated taking the ratio between the average of the output and input displacement frequency response. 

Two cases are considered. In a first step, the crystal is excited along its axial direction (input displacement of 1 mm in $z$ direction). This case represents an ideal configuration where the phononic crystal is used to attenuate perfect longitudinal motion. The second case considers the phononic crystal being mainly excited in the axial direction, however also slightly excited in its radial direction (input displacement of 1 mm in $z$ direction and 0.01~mm in $y$ direction). This study gives insight in the effect of slight misalignments of the vibration isolating crystals.

\subsection{Comparison of the experimental and numerical results.}
The measured and numerically calculated transfer functions for isotactic and syndiotactic arrangements are now compared. Two types of excitation are confronted: a perfect axial excitation is compared with a slightly off-axis excitation.
%excentric excitation in the direction of the main axis.

\begin{figure}[hbt!]
    \centering
\includegraphics[width=\textwidth]{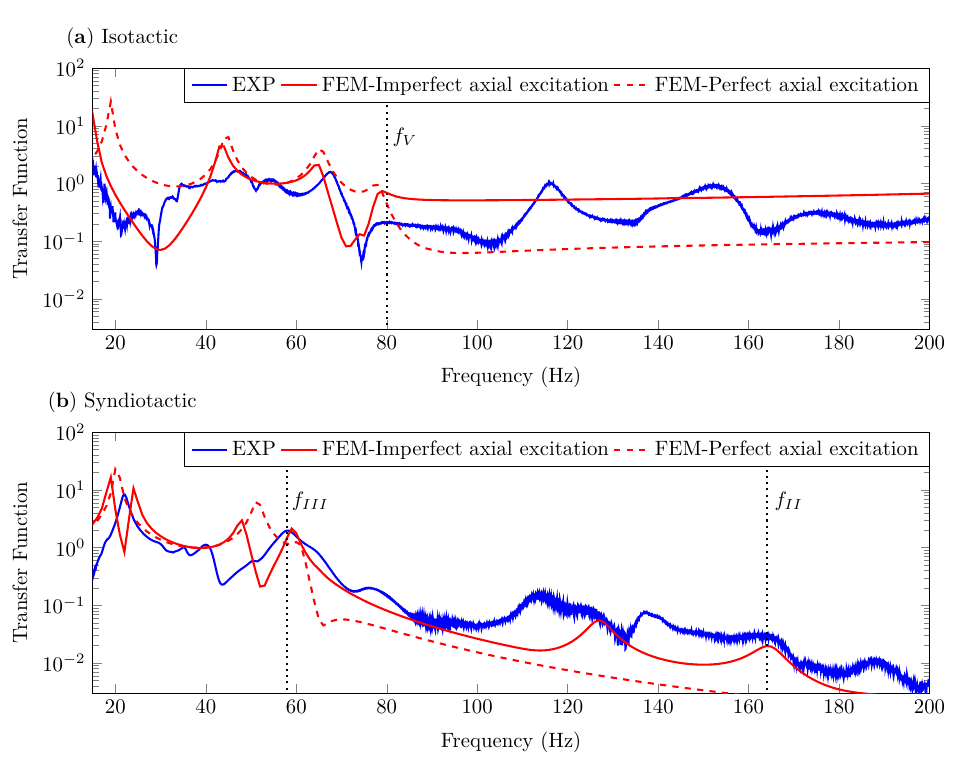}
    \caption{Transfer function of the isotactic (a) and syndiotactic (b) chiral phononic crystals.The measured transfer function is shown in blue, while the numerically calculated transfer function is shown in red under perfect axial excitation (dotted line) and imperfect axial excitation (continuous line).}
    \label{fig:TFSyndio}
\end{figure}

\begin{figure}[hbt!]
\centering
\includegraphics[width=\textwidth]{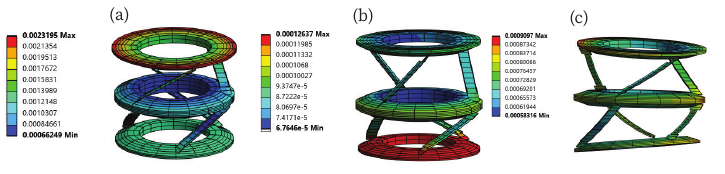}
        \caption{Deformation of the isotactic crystal under imperfect axial excitation showing the longitudinal resonance at 40 Hz (a) and the purely longitudinal motion under perfect excitation at 100 Hz (b) in opposition with the bending motion of the crystal under imperfect excitation (c).}\label{fig:DeformationsISO}
\end{figure}

As observed in Figure~\ref{fig:TFSyndio}~(a), the numerical and experimental transfer functions for the isotactic crystal show no clear attenuation over the entire frequency range, as no full bandgap was predicted by the unit cell analysis. The numerical and measured transfer functions show a comparable trend up to 80 Hz: a slightly attenuated frequency range up to 40 Hz, broken by a longitudinal resonance peak, shown in Figure~\ref{fig:DeformationsISO}~(a). This resonance is followed by a plateau up to 80 Hz. The crystal under perfect axial excitation shows a transfer function below 1 from 80 Hz onwards, corresponding to the attenuated range of the longitudinal mode V identified in the unit cell analysis. However, in this frequency range the crystal still deforms longitudinally as observed at 100 Hz Figure~\ref{fig:DeformationsISO}~(c): the longitudinal propagative mode I identified on the unit cell analysis contributes the most to the behavior of the crystal when subjected to external excitation. The attenuation is almost inexistent when looking at the transfer function under imperfect excitation, as well as for the measured transfer function. Indeed, the crystal shows a predominance of longitudinal motion up to 80 Hz mixed with flexural motion, illustrated in Figure~\ref{fig:DeformationsISO}~(b) for the 100 Hz mode. When the crystal is not perfectly excited along its axis, bending waves propagate, thereby reducing the efficiency of the partial longitudinal bandgap. From this finite structure analysis, we can see that attenuation at low frequencies can theoretically be observed within the isotactic crystal, however this attenuation is limited and not robust to excitation asymmetry, and therefore not suitable for practical applications.

\begin{figure}[hbt!]
        \centering      
\includegraphics[width=\textwidth]{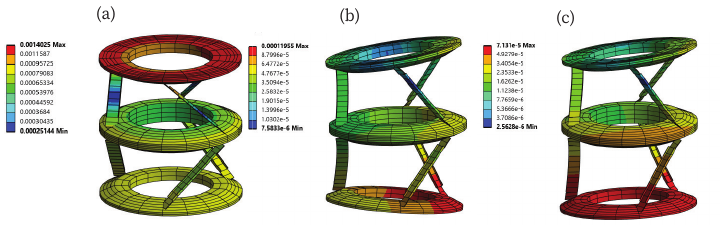}
        \caption{Deformation of the syndiotactic crystal under imperfect axial excitation showing the longitudinal resonance at 60 Hz (a) and bending resonances at 130 Hz (b) and 160 Hz (c).}\label{fig:DeformationsSYNDIO}
\end{figure}

The numerical and measured transfer functions of the syndiotactic crystal show a comparable trend over the entire range for the two excitation cases considered: high transmission up to 60 Hz, and subsequently decreasing from that starting frequency as shown in Figure~\ref{fig:TFSyndio}~(b).   
Three longitudinal resonance peaks are observed below the bandgap on the measured and calculated transfer functions. The resonance peaks are shifted in frequency in the numerical curves due to an approximated geometry of the struts in the models, and small geometric imperfections of the assembled crystal. However, the calculated bandgap starting frequency, with a deformation shown in Figure~\ref{fig:DeformationsSYNDIO}~(a), is in good agreement with the  measured value. The approximated FEM model, therefore, enables us to predict the attenuated frequency range with good accuracy, which makes its dimensioning sufficiently precise for demonstration purposes or for its implementation in applications. Furthermore, the starting frequency of the bandgap predicted for an infinite-size syndiotactic crystal is in agreement with the analysis for a finite-size. The dispersion curve calculated shows that the longitudinal motion of the crystal is significantly attenuated at frequencies above 60 Hz. As observed with the isotactic crystal, we see differences between the imperfect and perfect axial excitation cases. The transfer function of the crystal under imperfect axial excitation shows bending resonance peaks not observed in the perfect axial excitation case at 130 Hz Figure~\ref{fig:DeformationsSYNDIO}~(b) and 165 Hz Figure~\ref{fig:DeformationsSYNDIO}~(c) within the predicted longitudinal bandgap. This observation correlates with the dispersion curve analysis which exhibits a flexural mode propagating within the longitudinal bandgap. This analysis enables to demonstrate that the syndiotactic crystal is a robust solution even in the case of excitation asymmetry: the transfer function remains below 1 from 60 Hz. However, to allow its implementation in practical cases, a finite size analysis is necessary to avoid the presence of bending resonance at a predominant external excitation frequency. \\

\section{Conclusions}
This paper proposes a new design to build a large-scale chiral phononic crystal in order to demonstrate the abstract concepts of chirality and tacticity. For a fixed mass of the entire structure, the crystal's bandgap depends on the stiffness of the connectors (determined by their angle of inclination, thickness and shape) and on the inertial amplification of the masses (determined by the shape of the mass and on the positioning and inclination of the connectors). Based on these physical principles, the new geometry has been redesigned to limit the number of  assembly steps. The processes used to build the mechanical elements are conventional and do not require 3D-printed parts.  

A finite element model simplifying the geometry of the tilting connectors is proposed. Unit cell analysis and harmonic response analyses can be performed fast enough to do parametric studies and geometry optimizations. A parametric study highlights the relations between the new geometry and the crystal's tunability to achieve an effective vibration isolation frequency range while fixing the static stiffness and mass density within desired boundaries. The transfer function measured on the fabricated demonstrators shows comparable trends with the numerically calculated transfer function. As expected by the dispersion analysis, no clear bandgap is identified on the measured and numerically calculated under perfect and imperfect excitation transfer functions of the isotactic crystal. In the case of the syndiotactic crystal, good agreement between the longitudinal bandgap identified on the transfer function of the finite-size structure under perfect axial excitation and the dispersion curve is demonstrated. For non-perfectly axial excitation, bending modes resonance peaks appear within the attenuated frequency range as can be anticipated by the unit cell analysis. However, the resonance peaks remain at a sufficiently high attenuation level to confirm that syndiotactic crystal is a robust solution and that the bandgap can be predicted with an infinite-size analysis, even under imperfect excitation conditions.

\section*{Data availability}
Data will be made available on request.

\section*{Acknowledgements}
This work was done in the framework of the European Commission's Horizon Europe research and innovation programme under grant agreement No 101072415 and funded by the Swiss State Secretariat for Education, Research and Innovation (SERI), contract No 22.00225. Views and opinions expressed are however those of the authors only and do not necessarily reflect those of the European Union. The  European Union cannot be held responsible for them. Internal Funds KU Leuven are gratefully acknowledged for their support.

\section{Bibliography}
\label{sec:Bibliography}
%--------------------------------------------------%
\bibliographystyle{unsrtnat}

\bibliography{bib/PaperDemonstrator1DPhononicCrystal}

\end{document}